\documentclass[journal=jacsat,manuscript=article]{achemso}

\usepackage[version=3]{mhchem} 

\usepackage{appendix}
\usepackage{tikz}
\usetikzlibrary{shapes.geometric,arrows}
\tikzstyle{arrow} = [thick,->,>=stealth]
\usepackage{dcolumn}
\usepackage{tabularx}
\usepackage{booktabs}
\usepackage{hyperref}
\usepackage{dcolumn}
\usepackage{tabularx}
\usepackage{booktabs}
\usepackage{hyperref}
\usepackage{float}
\newcolumntype{Y}{>{\centering\arraybackslash}p{1.9cm}}
\newcolumntype{L}{>{\raggedright\arraybackslash}p{2.9cm}}
\newcolumntype{M}{>{\raggedright\arraybackslash}p{2.7cm}}
\newcolumntype{N}{>{\raggedright\arraybackslash}p{2.7cm}}
\newcolumntype{W}{>{\raggedright\arraybackslash}X}

\author{Mahmood Hasani}

\author{Hadis Salasi}

\author{Negar Ashari Astani}
\email{ashari@aut.ac.ir}
\affiliation[Unknown University]
{Department of Physics, Amirkabir University of Technology, Tehran, Iran}

\title[An \textsf{achemso} demo]
  {Quantum-Inspired Ising Machines for Quantum Chemistry Calculations}

\abbreviations{IR,NMR,UV}
\SectionNumbersOn
\begin{document}


\begin{abstract}
Four decades after Richard Feynman’s famous remark, we have reached a stage at which nature can be simulated quantum mechanically. Quantum simulation is among the most promising applications of quantum computing; however, like many quantum algorithms, it is severely constrained by noise in near-term hardware. Quantum-inspired algorithms provide an attractive alternative by avoiding the need for error-prone quantum devices.
In this study, we demonstrate that the coherent Ising machine and simulated bifurcation algorithms can accurately reproduce the electronic energy profiles of H$_{2}$ and H$_{2}$O, capturing their essential energetic features. Notably, we obtain computational times of 1.2 s and 2.4 s for the H$_{2}$ and H$_{2}$O profiles, respectively, representing a substantial speed-up compared to gate-based quantum computing approaches, which typically require at least 6 s to compute a single molecular geometry with comparable accuracy. These results highlight the potential of quantum-inspired approaches for scaling to larger molecular systems and for future applications in chemistry and materials science.
\end{abstract}

\section{Introduction}

Solving the Schrödinger equation accurately for atoms, molecules, and extended systems remains a central challenge due to the exponential scaling of electronic structure methods. Post-Hartree-Fock approaches such as Full Configuration Interaction (FCI, $O(N!)$) and Coupled Cluster (CC, $O(N^7)$) achieve chemical accuracy but are computationally intractable for large systems \cite{Whitfield2013,Schuch2013}.

Quantum computing offers a transformative alternative, as originally envisioned by Feynman \cite{Feynman1982}. Quantum Phase Estimation (QPE) promises FCI-level accuracy \cite{Aspuru2005} but requires fault-tolerant quantum computers (FTQC). Advances by Jones et al. \cite{Jones2012} established protocols (Programmable Ancilla Rotations (PARs), Phase Kickback, Fault-Tolerant Gate Approximation Sequences, etc.) for scalable ab initio simulations, while hybrid quantum-classical algorithms such as the Variational Quantum Eigensolver (VQE) \cite{Peruzzo2014,Farhi2014,Ion2019} emerged to address Noisy Intermediate-Scale Quantum (NISQ) era constraints. VQE reduces circuit depth by combining parameterized quantum circuits with classical optimization, but faces challenges including ansatz design \cite{Bartlett1989}, barren plateaus \cite{MacClean2018}, and measurement overhead \cite{Areview2019}.

Numerous improvements, such as UCCSD \cite{Pal1984,Taube2006,Sur2008,Cooper2010,Harsha2018,Evangelista2019,Shen2017}, ADAPT-VQE \cite{Grimsley2019}, k-UpCCGSD \cite{Lee2018}, Qubit-ADAPT-VQE \cite{Tang2021}, and Quantum Subspace Expansion (QSE) \cite{Yoshioka2022}, have been developed, along with hardware-efficient architectures and quantum machine learning extensions \cite{Mangini2023}. Real hardware implementation of these algorithms based on photonics \cite{Peruzzo2014, Wang2015}, superconducting qubits \cite{Omalley2016,Kandala2017,Kandala2019,colless2017}, and trapped ions \cite{Hempel2018,Nam2020}, highlighting both promise and limitations. Beyond gate-based approaches, real-space, real-time simulations offer direct dynamical insights at high computational cost \cite{Ward2009,Kassal2011,Welch2014,WhitfieldSpin2013,Lu2011,Childs2022}.

Adiabatic quantum computing and quantum annealing also provide alternative formulations. Although limited by hardware connectivity, the mapping of the electronic structure to the Ising or Quadratic unconstrained binary optimization (QUBO) Hamiltonians has been demonstrated on D-Wave devices \cite{Babbush2015,Xia2017,Streif2019}. Hybrid strategies, such as Qubit Coupled Cluster (QCC) \cite{Genin2019} and Quantum Annealer Eigensolver (QAE) \cite{Teplukhin2020}, extend these approaches to intermediate-size molecular systems. Collectively, these efforts underscore both the potential and current limitations of quantum algorithms for chemistry.

Recently, Quantum Annealing-Inspired Algorithms (QAIAs) have emerged as scalable, high-performance alternatives. While not inherently quantum, they leverage principles of quantum annealing to achieve more than $98\%$ accuracy in Ising ground-state estimation. Two leading approaches are Coherent Ising Machines (CIMs), based on networks of quantum optical parametric oscillators \cite{Marandi2013,Marandi2014,mcmahon2016fully,inagaki2016coherent,Yamamura2017,Yamamoto2017,Lvovsky2019,Reifenstein2021,Hamerly2019,Honjo2021}, and Simulated Bifurcation (SB) algorithms, which exploit nonlinear bifurcation dynamics of Kerr-nonlinearity for efficient optimization \cite{Goto2019,Goto2021,Wang2023}. CIMs benefit from all-to-all optical coupling and large-scale operation at room temperature, while SB algorithms offer computationally efficient variants successfully applied to molecular Hamiltonians.

In this work, we employ a hybrid quantum-inspired algorithm that integrates Coherent Ising Machine (CIM) and Simulated Bifurcation (SB) methods with a refined steepest-descent post-processing scheme to reconstruct molecular energy landscapes. By comparing the performance of different CIM and SB implementations within this hybrid framework, we demonstrate the applicability of QAIAs as accurate and scalable tools for quantum chemistry. Notably, our approach achieves a dramatic acceleration in computational throughput, resolving the complete energy profiles for $H_2$ and $H_2O$ in just 1.2 s and 2.4 s, respectively, while maintaining an accuracy comparable to standard quantum approaches. This efficiency stands in sharp contrast to current quantum hardware, which typically requires upwards of 6 s to compute a single energy point with similar error margins. Consequently, our results suggest that QAIAs offer a powerful, immediate alternative for bypassing current hardware bottlenecks to simulate complex chemical systems.

 Building upon the fermion-to-Ising mapping of these systems (Subsec. \ref{sec:mappings}), we employ variants of the SB and CIM algorithms as Ising solvers, integrated with a steepest descent scheme as a greedy post-processing procedure (Subsec. \ref{sec:isingmachines}), to obtain their ground-state energy profiles (Sec.~\ref{sec:methods}). We then analyze and benchmark the results against exact methods, Complete Active Space Configuration Interaction (CASCI), and Hartree–Fock (HF) (Sec.~\ref{sec:results}). Finally, in Sec.~\ref{sec:conclusion}, we discuss the broader implications of this approach and its potential extension to larger and more complex molecular systems.

\section{\label{sec:methods}Methodology}

\subsection{Mapping the Molecular Hamiltonian into Ising Hamiltonian \label{sec:mappings}}

Mapping a molecular Hamiltonian into an Ising-type Hamiltonian is a multistep procedure that connects {ab initio} quantum chemistry with quantum optimization and quantum annealing methods. A detailed description of this framework is provided by Xia et al. \cite{Xia2017}. The process begins with the construction of molecular orbitals, typically from a chosen atomic orbital basis, such as $STO-6G$ or larger basis sets, depending on the required accuracy. Spin orbitals are then obtained by combining spatial orbitals with spin functions while also accounting for the overlap integrals between atomic orbitals.  

With these orbitals defined, the molecular electronic Hamiltonian can be expressed in second-quantized form as
\begin{equation}
    H = \sum_{pq} h_{pq}\, a_p^\dagger a_q + \frac{1}{2} \sum_{pqrs} h_{pqrs}\, a_p^\dagger a_q^\dagger a_r a_s,
\end{equation}
where $a_p^\dagger$ and $a_q$ denote fermionic creation and annihilation operators, and the coefficients $h_{pq}$ and $h_{pqrs}$ are one- and two-electron integrals representing kinetic energy, nuclear attraction, and electron-electron repulsion.  

Since quantum hardware does not directly support fermionic operators, the Hamiltonian is mapped to qubit operators through transformations such as the Jordan-Wigner or Bravyi-Kitaev transformations. These mappings convert the fermionic operators into tensor products of Pauli matrices ($\sigma_x, \sigma_y, \sigma_z$), yielding a qubit Hamiltonian. Furthermore, molecular symmetries and conservation laws can be leveraged to reduce the number of active qubits, which is crucial for the efficient implementation of quantum devices.  

The resulting qubit Hamiltonian takes the form of a weighted sum of Pauli strings. To adapt this representation to quantum annealers, which natively operate with Ising-type models supporting only $2$-local interactions, higher-order terms are systematically reduced using auxiliary variables and penalty constraints. This procedure yields the standard Ising Hamiltonian representation:
\begin{equation}
    H_{\text{Ising}} = \sum_i h_i z_i + \sum_{i<j} J_{ij} z_i z_j, 
    \quad z_i \in \{-1,+1\},
\end{equation}
where $h_i$ are local fields and $J_{ij}$ represent pairwise couplings.  

This general workflow applies to arbitrary molecular systems. In this paper, we explicitly demonstrate its implementation for two representative molecules: $H_{2}$ and $H_{2}O$. The complete derivations are available in the {Supplementary Information} (SI).

\subsection{Ising Machines \label{sec:isingmachines}}

Several approaches have been proposed for sampling data to feed into Ising solvers for molecular electronic structure calculations. Previous studies have employed D-Wave annealers for this purpose \cite{Babbush2015,Xia2017,Streif2019}. In this work, we employ quantum-inspired Ising solvers, specifically the CIM and the SB algorithm, which offer distinct advantages over D-Wave–based quantum annealing approaches. Both CIMs and SBs have demonstrated high efficiency in determining the minimum energy configurations of the Ising Hamiltonian, often achieving superior success probabilities. The operational principles and algorithmic structures of these solvers are described in detail in the following section.

From a physical standpoint, the operation of CIMs can be understood in terms of oscillator dynamics: each oscillator undergoes a parametric gain that amplifies the field once the pump exceeds the threshold, while nonlinear loss (saturation) constrains the amplitude to one of two stable phase states \cite{Wang2013}. The couplings between oscillators are implemented by modulating the pump or injection signals according to the Ising coupling matrix $J_{ij}$, thereby encoding the global Ising energy into the gain–loss landscape of the entire network. Quantum and classical noise seed the initial phases, and the ensuing nonlinear dynamics guide the system toward low-energy phase configurations. Thus, CIMs exploit the physics of parametric oscillation and nonlinear bifurcation to identify the ground-state configuration.

Previous studies have proposed a range of quantum-inspired algorithms based on CIMs that address the Ising problem through the Heisenberg–Langevin formalism, most notably the Chaotic Amplitude Control (CAC), Chaotic Feedback Control (CFC), and Separated Feedback Control (SFC) algorithms \cite{Reifenstein2021,Leleu2021}.

To numerically emulate the physical behavior of CIMs, the Simulated Coherent Ising Machine (SimCIM) framework is often employed. In this approach, the nonlinear oscillator dynamics are modeled through stochastic differential equations that evolve hyperparameters as dynamical variables under the influence of gain and coupling terms.

Within the SimCIM framework, we evaluate three distinct dynamical formulations to determine the most effective solver for our molecular landscapes. The first variant, known as Chaotic Amplitude Control (CAC), regulates the system's trajectory by correcting errors based on the spin amplitude. The time evolution of the spin variable $x_i$ and the error variable $e_i$ in the CAC variant is governed by the following stochastic differential equations:
\begin{equation} \frac{dx_i}{dt} = -x_i^3 + (p-1)x_i + e_i \sum_j \zeta J_{ij} x_j, \end{equation}
\begin{equation}\label{errorCAC} \frac{de_i}{dt} = -\beta e_i (x_i^2 - \alpha). \end{equation}
where 
$p$,$\alpha$
,$\beta$ , and $\zeta$ denote the gain parameter, the target amplitude, the rate of change of the error variables, and the coupling strength, respectively. Also, 
$J_{ij}$ represents the Ising coupling matrix.
To explore the impact of the error mechanism on convergence, we consider a second variant: Chaotic Feedback Control (CFC). In contrast to CAC, which utilizes the local spin amplitude 
$x_i$ for error correction (Eq. \ref{errorCAC}), the CFC algorithm couples the error variable to the local mean field $z_i$. The time evolution for CFC is described as follows:
\begin{equation}\label{z_i_cfc} z_i = -e_i \sum_j \zeta J_{ij} x_j, \end{equation} \begin{equation} \frac{dx_i}{dt} = -x_i^3 + (p-1)x_i - z_i, \end{equation} \begin{equation}\label{errorCFC} \frac{de_i}{dt} = -\beta e_i (z_i^2 - \alpha). \end{equation}
As observed in Eq. (\ref{errorCFC}), the error dynamics are driven by the auxiliary field 
$z_i$, providing an alternative feedback pathway compared to the CAC model.
Finally, we investigate the Simulated Feedback Control (SFC) variant. This formulation differs from the previous two by introducing a hyperbolic tangent nonlinearity to strictly govern the transition from the continuous (soft-spin) to the binary (discrete-spin) regime. The time evolution is written as:
\begin{equation}\label{z_i_sfc} z_i = -\sum_j \zeta J_{ij} x_j, \end{equation} \begin{equation} \frac{dx_i}{dt} = -x_i^3 + (p-1)x_i - \tanh(cz_i) - k(z_{i} - e_{i}), \end{equation} \begin{equation}\label{errorSFC} \frac{de_i}{dt} = -\beta (e_i - z_i). \end{equation}
In these equations, the parameter 
$c$
 governs the steepness of the nonlinearity, while the error evolution in Eq. (\ref{errorSFC}) adopts a linear tracking of the mean field difference, distinct from the multiplicative error terms in CAC and CFC.

In summary, CIMs leverage the nonlinear dynamics of coupled parametric oscillators to encode and minimize the Ising energy, offering a physically motivated route to ground-state identification. 

Beyond CIM variants, the discrete Simulated Bifurcation (dSB) algorithm provides a highly efficient quantum-inspired framework for solving Ising problems. Rooted in the bifurcation theorem and the adiabatic evolution of nonlinear Hamiltonian systems \cite{Goto2016}, the dSB algorithm is particularly noted for its parallel processing capabilities and its ability to rapidly locate local minima near the ground state \cite{Goto2019,Goto2021}. Unlike the standard ballistic formulation, the dSB variant enhances optimization performance by discretizing the variables within the interaction term. The time evolution of the spins is governed by:
\begin{equation}\label{SBposition}
\frac{dx_i}{dt} = a_0 p_i
\end{equation}
\begin{equation}\label{SBmomentum}
\frac{dp_i}{dt} = -(a_0 - a(t))x_i + c_0 \sum_{j=1}^{N} J_{ij} \text{sgn}(x_j),
\end{equation}
where $x_i$ and $p_i$ denote the position and momentum of the $i$-th Kerr nonlinear parametric oscillator, respectively. The parameter $a_0$ represents the positive detuning frequency, $a(t)$ is the time-dependent pumping amplitude, and $c_0$ denotes the coupling strength. Crucially, as shown in Eq. (\ref{SBmomentum}), the dSB algorithm employs the sign of the position variable, $sgn(x_j)$, in the coupling term rather than the continuous variable $x_j$. This discretization enables the algorithm to explore the solution space more aggressively at the onset of iterations, facilitating faster convergence and improved escape from local minima.
To constrain the dynamics, the algorithm incorporates perfectly inelastic walls at $x_i=±1$. If a variable $x_i$ exceeds these bounds, it is immediately reset to the boundary value, and its momentum $p_i$ is set to zero. This mechanism ensures that the system stabilizes into discrete binary states as the pumping amplitude $a(t)$ increases. We focus exclusively on this dSB implementation as it represents the most robust and powerful variant within the simulated bifurcation framework.

To further improve the solutions obtained from the Ising solvers, namely, the CIMs and SB, we apply a post-processing technique based on a greedy algorithm, specifically the steepest descent method provided by the \texttt{SteepestDescentSolver} in D-Wave's Ocean software \cite{Teukolsky1992}. This method iteratively refines the spin configurations returned by the quantum processing unit (QPU), aiming to minimize the energy of the system and identify lower-energy configurations. In essence, once a local ground state is obtained from the Ising solver, the algorithm explores its neighboring configurations to ensure that the best possible local minimum is achieved.

In the steepest-descent method, the algorithm evaluates the energy change associated with flipping each spin. The energy change resulting from flipping spin \( s_i \) is given by:
\begin{equation}
\Delta E_i = 2 s_i \left( h_i + \sum_{j \neq i} J_{ij} s_j \right).
\end{equation}
At each iteration, the algorithm selects the spin flip that produces the most negative \( \Delta E_i \), thereby maximizing the reduction in the system's energy. This process continues until no spin flip yields a negative energy change, indicating that a local minimum has been reached.

The greedy algorithm offers a computationally efficient and straightforward means to refine QPU solutions; thus, its performance can be limited by convergence to local minima in complex Ising energy landscapes. To address this, multiple QPU samples are employed as independent initial states. In that, sampling with CIMs and SB involves generating independent spin configurations from the dynamical evolution of the CIMs and SB algorithms under varying initial conditions or noise realizations. This stochastic sampling strategy enhances exploration of the energy landscape, increasing the likelihood of locating low-energy or ground-state configurations.

\begin{figure}[h]
    \centering
    \includegraphics[width=1.0\textwidth]{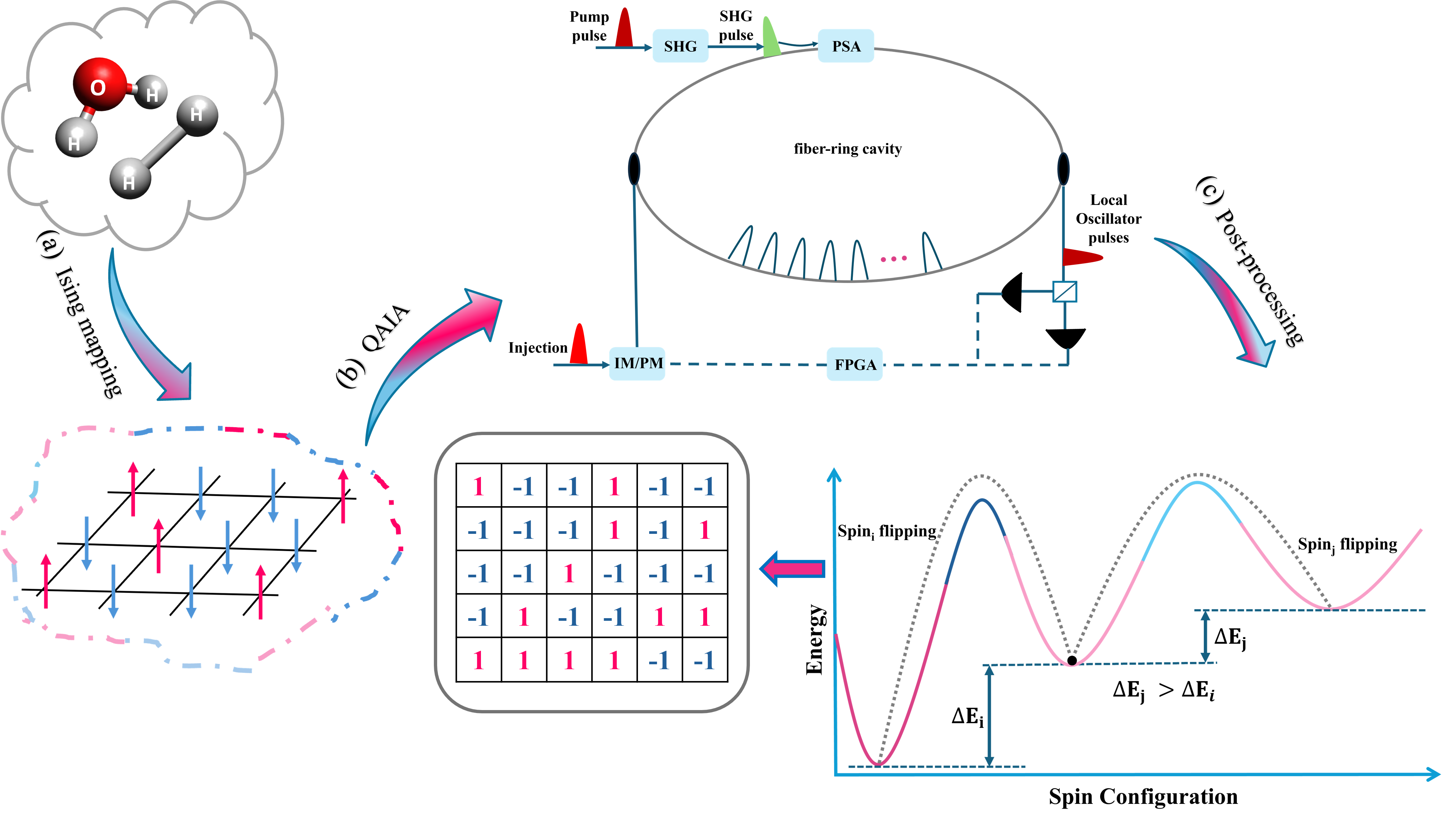}
    \caption{Schematic presentation of the hybrid quantum–inspired algorithm used to determine the ground-state energy profiles of \(H_2\) and \(H_2O\). (a) Conversion of the molecular Hamiltonian into Ising form. (b) Feeding sampled data into quantum-inspired algorithms. (c) Applying the steepest descent method for result refinement.}
    \label{fig:hybrid_algorithm}
\end{figure}

\begin{figure}[h]
    \centering
    \includegraphics[width=0.65\textwidth]{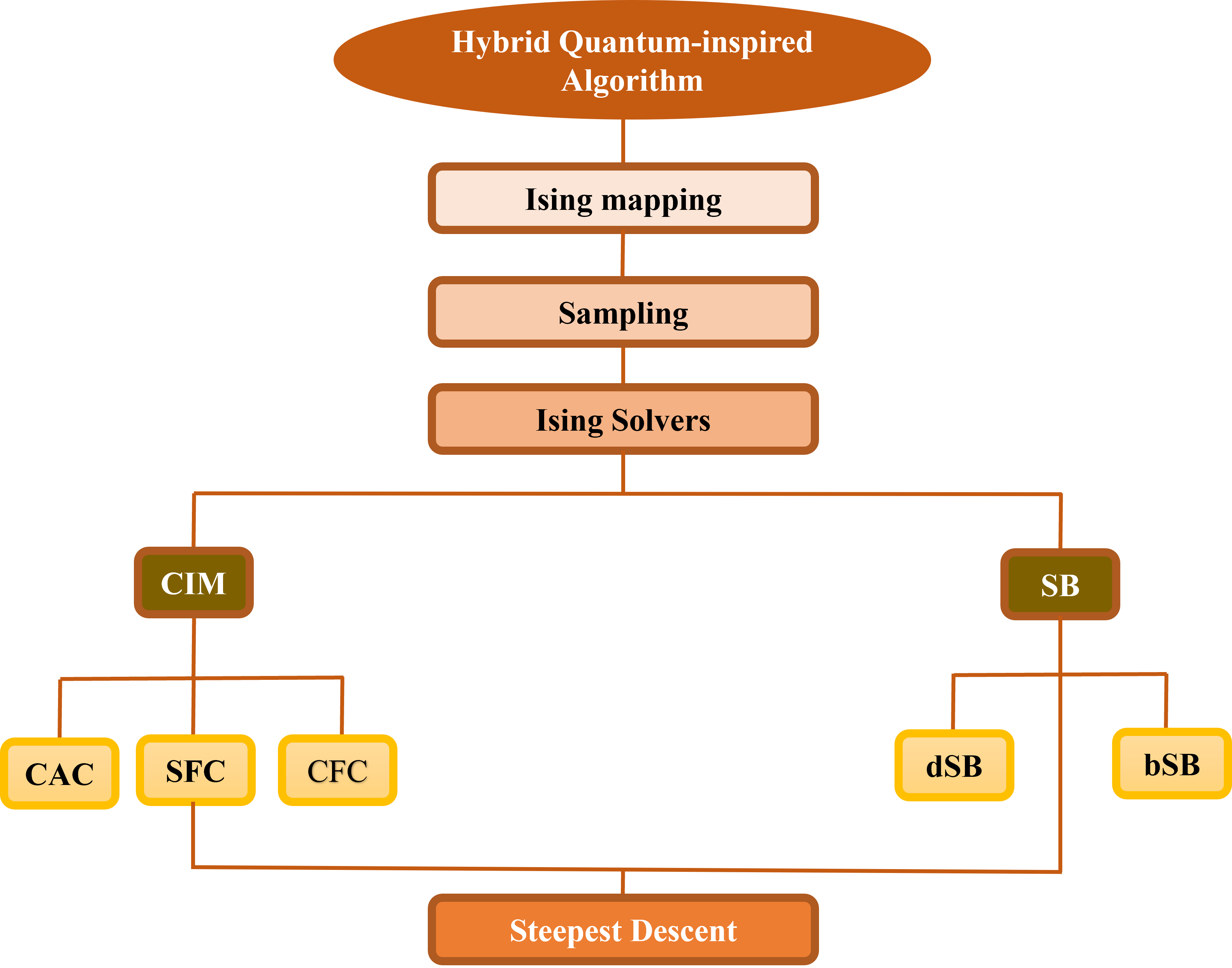}
    \caption{Schematic of the hybrid algorithm employed in this study, extending and refining the framework originally introduced by Yudong Cao et. al. \cite{Areview2019}.}
    \label{fig:electronic_structure_progress}
\end{figure}

To illustrate the overall methodology, Fig.~\ref{fig:hybrid_algorithm} presents a schematic of the hybrid algorithm, outlining the key steps from Hamiltonian mapping to final energy determination using the combined quantum-inspired and classical approach. In addition, Fig.~\ref{fig:electronic_structure_progress} provides a flowchart representation of the procedure discussed above. This figure builds upon the framework introduced by Yudong Cao et. al. \cite{Areview2019}, highlighting the progression of computational strategies in quantum chemistry and situating our hybrid approach within this broader context.

\section{Results}\label{sec:results}
This section presents the numerical results obtained by applying various discussed quantum-inspired Ising solver algorithms, including variants of the CIM and SB algorithms, to compute the ground-state energies of $H_2$ and $H_2O$ molecules. The performance of these algorithms is evaluated based on the chosen hyperparameters and computational strategies, using $100$ Ising configurations generated by the CIM and SB algorithms. A steepest-descent post-processing step is applied to further refine the results, as shown in Figures \ref{fig:2}, \ref{fig:3}, and \ref{fig:4}.

To assess the accuracy of our algorithms, we compare their results with those obtained from standard electronic structure methods, namely the Hartree-Fock (HF) and Complete Active Space Configuration Interaction (CASCI) methods. The HF method represents the molecular wavefunction as a single Slater determinant and provides the variationally optimal solution within the mean-field approximation, whereas the CASCI method yields the exact solution for the chosen basis set and active space and becomes equivalent to the full configuration interaction (FCI) method when all spin-orbitals are included in the active space \cite{Roos1987,Copenhaver2021}. These reference methods, therefore, serve as benchmarks to evaluate the performance and accuracy of the proposed algorithms.

We introduce the full set of hyperparameters for the SB and CIM variants, outlining practical strategies for tuning them to achieve accurate ground-state energy estimations in the {Supplementary Information}. The main results of this study are organized into two subsections, focusing on single-shot (Subsec.\ref{sec:singleshot}) and multi-shot (Subsec.\ref{sec:multishot}) sampling methods utilizing QAIAs. We define the single-shot approach as a solitary execution of the algorithm, whereas the multi-shot approach leverages GPU acceleration to generate a large ensemble of Ising configuration samples.

\subsection{Single-shot\label{sec:singleshot}}
\begin{figure*}[htbp]
  \centering
  \begin{tabular}{cc}
    \includegraphics[width=0.47\textwidth]{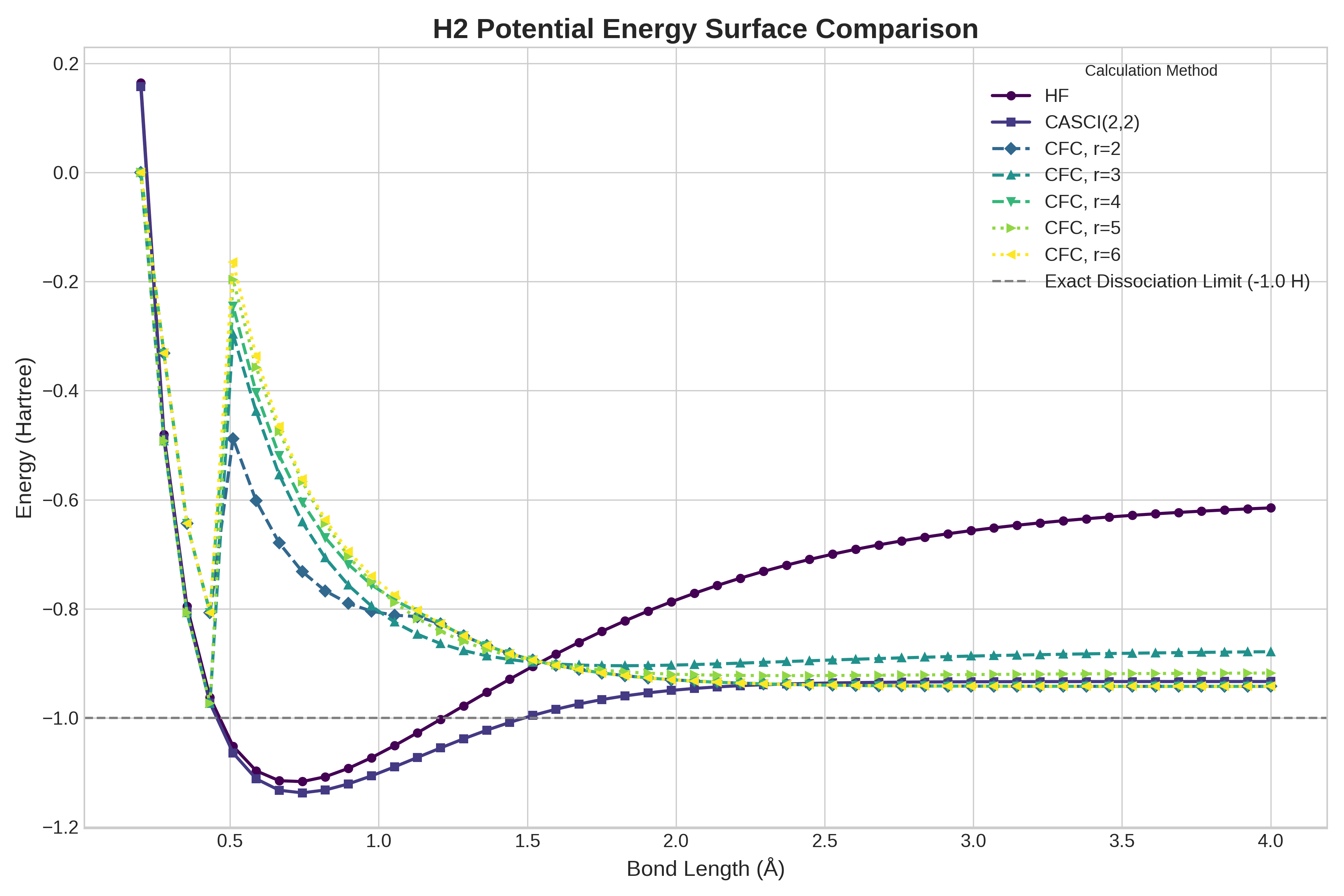} &
    \includegraphics[width=0.47\textwidth]{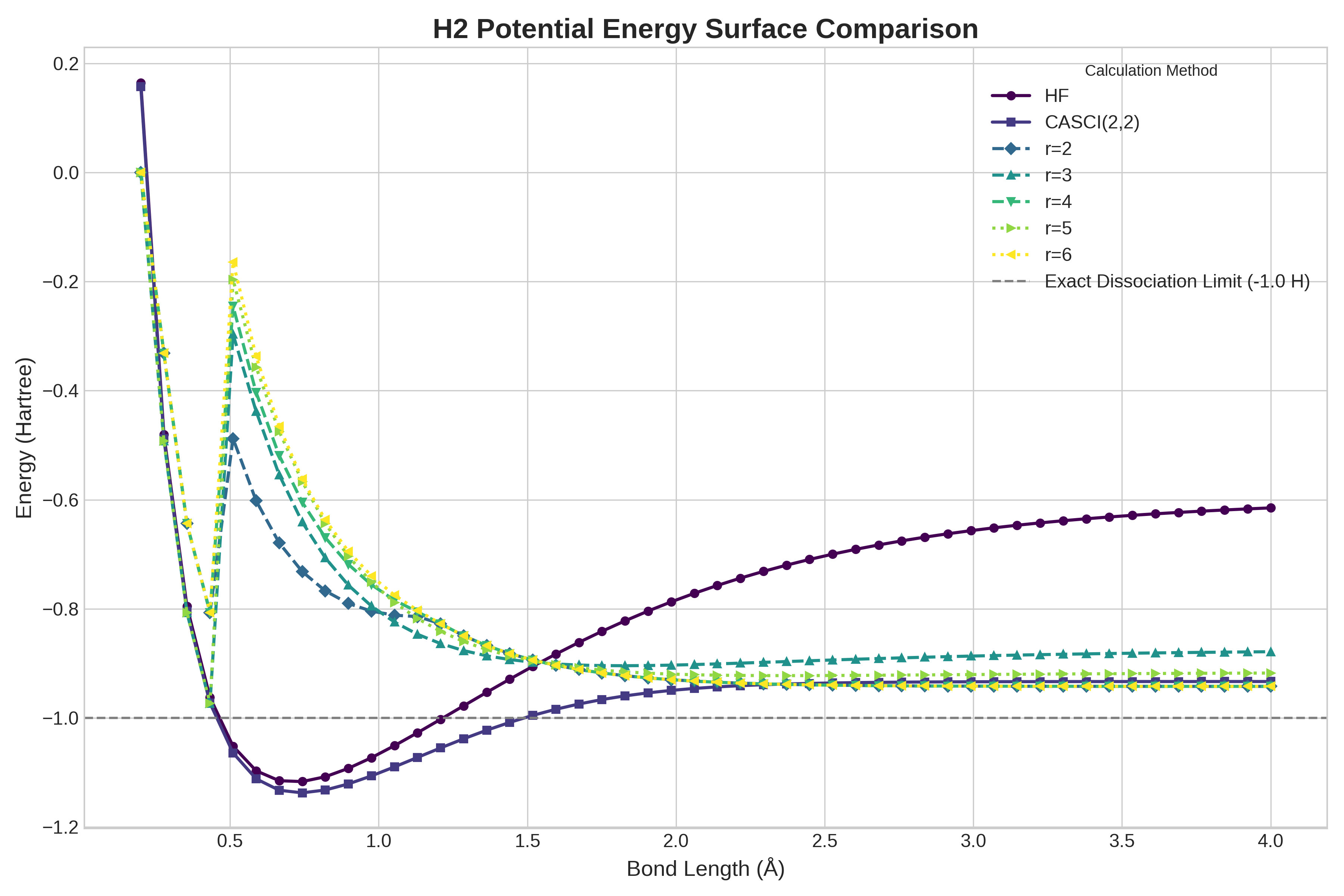} \\[1ex]
    \includegraphics[width=0.47\textwidth]{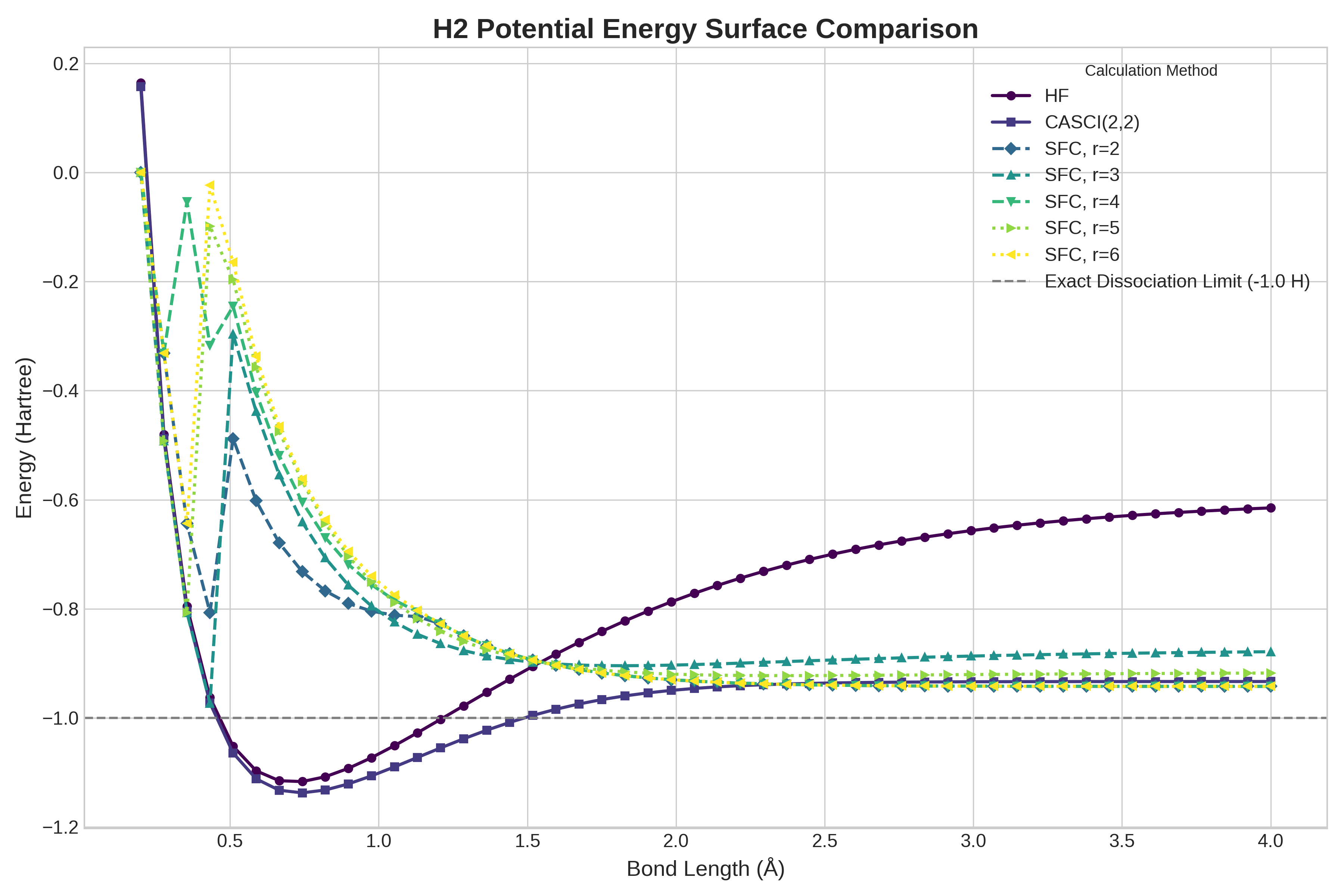} &
    \includegraphics[width=0.47\textwidth]{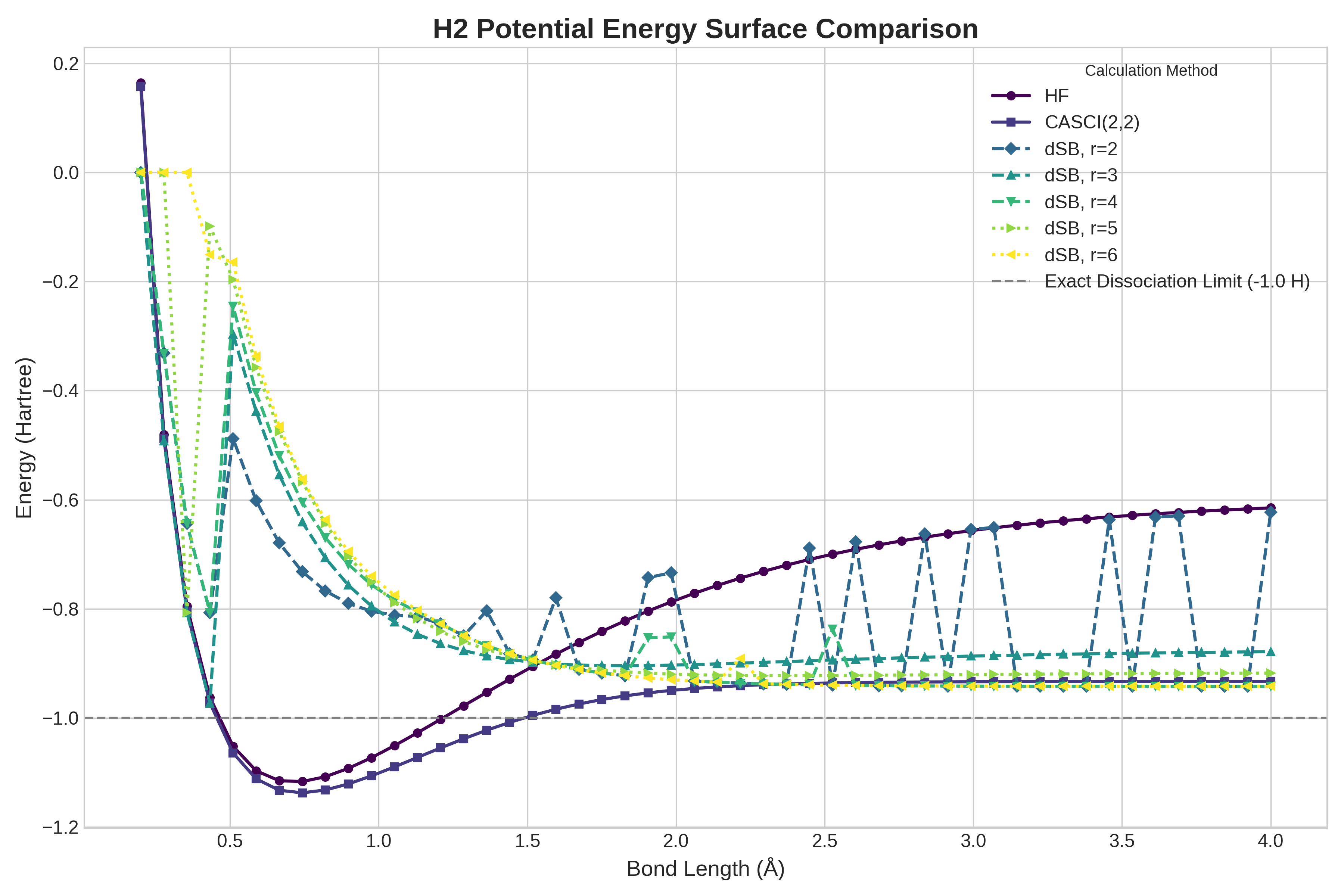} \\
  \end{tabular}
  \caption{Calculated ground-state energy of the hydrogen molecule as a function of internuclear distance using four quantum-inspired algorithms based on a single computational sample. The subfigures correspond to: (a) Separated Feedback Control (SFC), (b) Chaotic Feedback Control (CFC), (c) Discrete Simulated Bifurcation (dSB), and (d) Chaotic Amplitude Control (CAC). In each plot, the different colored lines represent variations in the key algorithm parameter $r$ (ranging from $r = 2$ to $r = 6$), illustrating its influence on the energy estimation.}
  \label{fig:2}
\end{figure*}

Fig. \ref{fig:2} illustrates the calculated ground-state energy of the hydrogen molecule as a function of its internuclear bond length. The results are presented for four quantum-inspired algorithms: SFC, CFC, dSB, and CAC. Each subfigure displays the corresponding energy profile, where the different colored lines represent variations in $r$, ranging from $r = 2$ to $r = 6$, a variational parameter introduced in the original paper \cite{Xia2017}.

Across all four algorithms, a consistent trend characteristic of molecular dissociation curves is observed. As the internuclear bond length increases from very short distances, the energy initially decreases, reaches a minimum, and then gradually increases, eventually approaching a plateau at larger bond lengths. This minimum energy corresponds to the equilibrium bond length and the ground-state energy of the molecule. The parameter $r$ influences both the accuracy and convergence of the energy estimation, with certain values of $r$ exhibiting closer agreement with the expected energy profile, particularly near the equilibrium bond length. In the XBK method, the energy plotted as a function of $r$ represents the collective ground-state energy of the molecule, as the transformation encodes the full molecular Hamiltonian into an expanded qubit space, capturing all interatomic interactions simultaneously rather than individual bond energies \cite{Xia2017,Copenhaver2021}.

\subsection{Multi-shot\label{sec:multishot}}

\begin{figure}
    \centering
    \includegraphics[width=0.9\columnwidth]{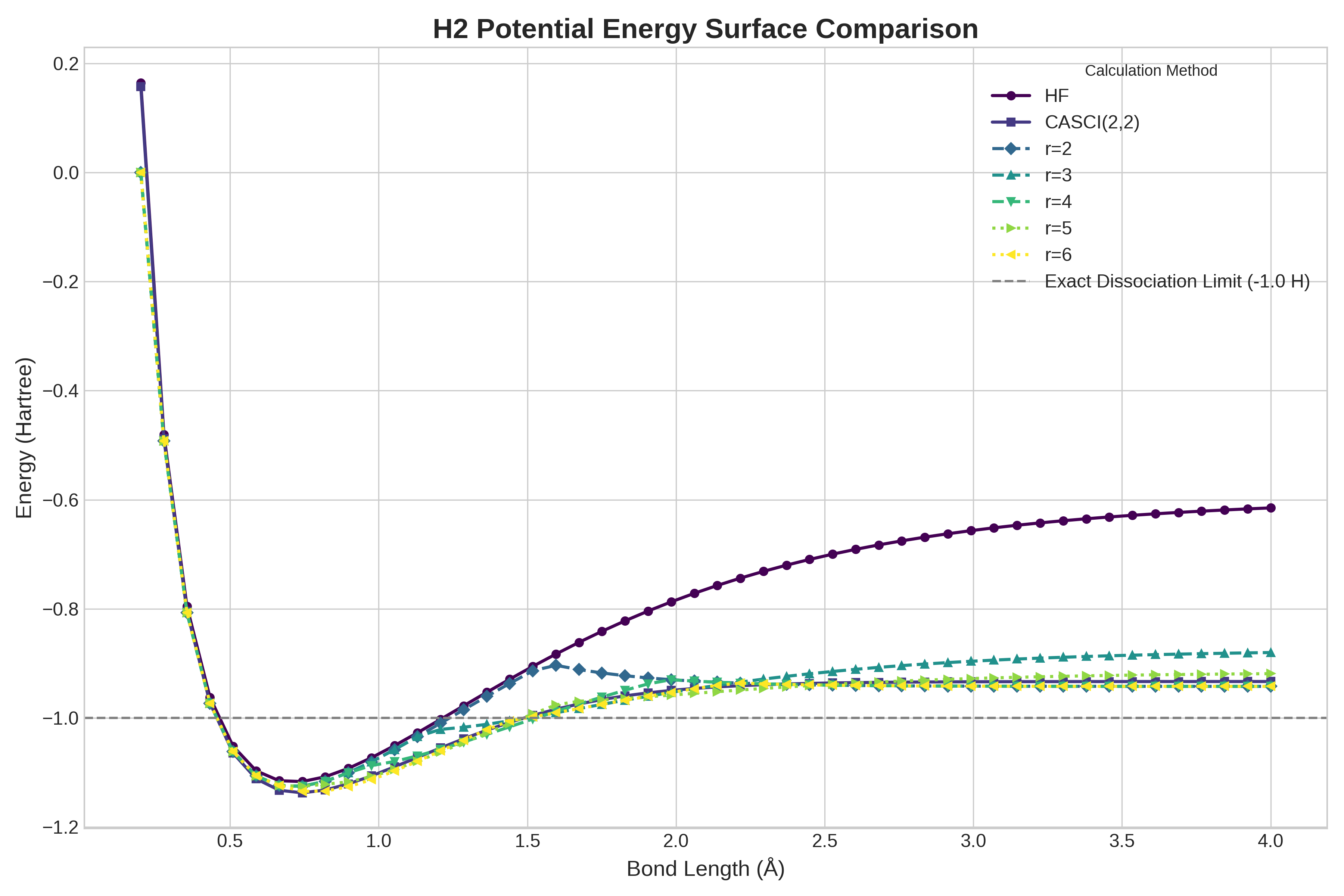}
    \caption{Performance of solving the hydrogen Hamiltonian for $r = 2$ and $r = 6$. Two experimental approaches are presented: increasing the number of Ising machine samples up to $100$, which allows the algorithms to reach the true global minimum, and using a single-shot execution of the Ising machines combined with a steepest descent algorithm, which achieves the same result.}
    \label{fig:3}
\end{figure}

Fig. \ref{fig:3} illustrates the performance of the algorithms in solving the hydrogen molecular Hamiltonian over the bond-length range $r=2–6$. The dashed line labeled “Exact Dissociation Limit” denotes the molecular energy when the constituent atoms are infinitely separated, corresponding to a completely broken chemical bond with no residual interaction. The figure compares two approaches for identifying the true global minimum. Based on our numerical tests across various CIM and SB algorithm variants, the CFC variant consistently required the fewest samples to achieve the exact electronic ground-state energy (table \ref{tab:sample}). Therefore, in subsequent analyses, multishot results for both molecules are reported exclusively using the CFC algorithm.

\begin{table}[h!]
\centering
\caption{Multi-Shot experiment for Ising Machines. Our benchmarking shows that the CFC variant of CIM can generate more accurate samples than others.}
\begin{tabular}{|c|c|c|}
\hline
{Ising Machine} & {samples}  \\ 
\hline
CFC & 100 \\ 
\hline
CAC & 700 \\ 
\hline
SFC & 500  \\ 
\hline
dSB & 500 \\ 
\hline
\end{tabular}
\label{tab:sample}
\end{table}

Firstly, the Fig. \ref{fig:3} shows that by increasing the number of samples for the Ising machines up to $100$ enables the algorithms to reach the true global minimum consistently. That is the dynamical equations of the system are solved $100$ times, each with a different random seed. This result suggests that using a sufficiently large number of computational samples enables a more comprehensive exploration of the solution space, thereby increasing the likelihood of accurately identifying the ground-state energy.

The Fig. \ref{fig:3} further suggests that even a single measurement, when supplemented with classical optimization techniques such as steepest descent, can guide the system toward the correct solution. This approach has the potential to reduce the computational overhead associated with multiple sampling iterations. Together, these results highlight the robustness of quantum-inspired algorithms, demonstrating their flexibility in adapting to different computational strategies and resource constraints while maintaining high accuracy.

It is noteworthy that GPU-based sampling of a larger number of Ising configurations yields comparable Time-to-Solution (TTS) values owing to inherent parallelization. However, using fewer samples offers a more resource-efficient implementation with reduced memory demand, where the CFC variant demonstrates superior performance under limited sampling conditions (see Table \ref{tab:tts_comparison}). Consequently, all subsequent experiments in this work employ the CFC variant of the CIM. Full description of Table \ref{tab:tts_comparison} will come later in the manuscript.

\begin{figure}
    \centering
    \includegraphics[width=1\columnwidth]{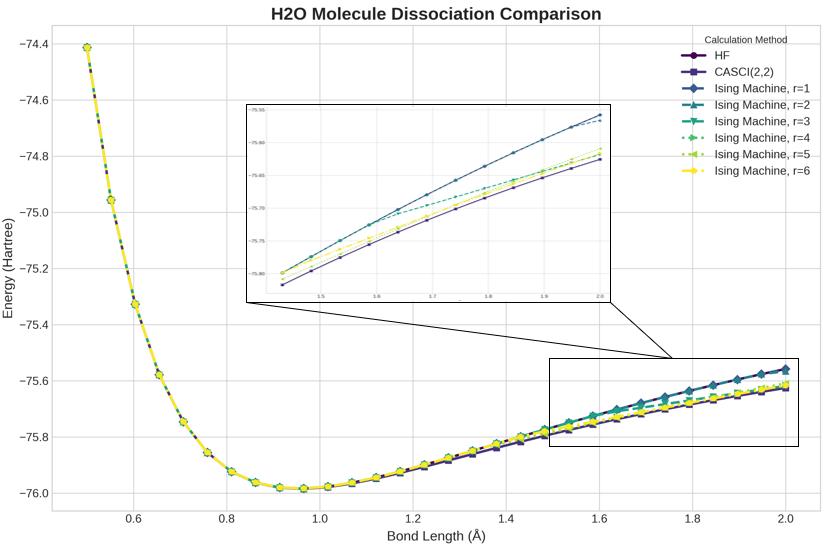}
    \caption{Energy profile of the water molecule computed using the CFC variant of the CIM, selected for its superior performance. The energy was evaluated over the bond-length range $r=[2,6]$.}
    \label{fig:4}
\end{figure}

Fig. \ref{fig:4} presents the energy profile of the water molecule, plotted for the same range of $r$ values ($r = 2$ to $r = 6$) as used for the hydrogen molecule. Although determining the exact numerical values and the detailed shape of the curve is more challenging due to the larger size and more intricate electronic structure of $H_2O$ compared to $H_2$, this figure demonstrates the applicability and scalability of quantum-inspired algorithms to more complex molecular systems. 

The parameter $r$ continues to influence the accuracy and behavior of the energy calculation, significantly consistent with its effect in the simpler $H_2$ system. Showcasing the energy profile of the water molecule serves as proof of principle, establishing the versatility of the proposed algorithm beyond simple diatomic systems.

Table \ref{tab:tts_comparison} summarizes the measured and estimated per-evaluation TTS for small-molecule ground-state energy calculations across gate-model VQE, annealer-based XBK/QCC, and quantum-inspired Ising-machine simulations. Two gate-model experiments on real hardware illustrate the wide range in end-to-end runtime: a superconducting IBMQ Manila run required approximately $334 \pm 40$ s per energy evaluation, while an ion-trap experiment on AQT Marmot reported much longer durations of $13297 \pm 625$ s, both including queueing, compilation, execution, and readout overheads \cite{Bentellis2023}. In addition, a quite recent work on superconductor quantum computer for finding the ground state of molecule reported that the time to find a precise ground state took about 1.5 hours.

Annealer-based approaches (XBK/QCC) executed on D-Wave yield faster single-evaluation times in the 1-9 seconds \cite{Copenhaver2021, Teplukhin2021}. However, they include several computationally expensive components, such as quadratization, embedding, and multiple qbsolv \cite{qbsolv} subcalls (qbsolv is a Python library utilized as a decomposition solver), and both classical and quantum contributions scale rapidly with system size. The estimated runtime corresponds to the end-to-end wall-clock time on the D-Wave quantum annealer, including fixed programming overhead and per-shot readout overhead. The actual quantum annealing time per run is on the order of microseconds and contributes only a negligible fraction to the total runtime.

By contrast, the quantum-inspired Ising-machine simulations proposed in this work, achieve short wall-clock times per energy evaluation, $1-3$ s on GPUs. In particular, total potential energy computation for H$_{2}$ required only $1.2$ s per bond-length parameter $r$ (Fig. \ref{fig:3}), while for H$_{2}$O the simulator required $2.4$ s per parameter $r$ (Fig. \ref{fig:4}). Crucially, this highlights a distinct practical advantage of the GPU-based Ising approach by bypassing the queueing latencies and programming overheads inherent to current cloud-based QPUs; we achieve superior computational throughput and immediate scalability.

We note two important caveats in benchmark studies. First, the reported simulator timings reflect only the algorithmic core and therefore exclude QPU programming latencies that would increase real-device TTS. Another problem is the observed scaling behavior may differ for larger molecular systems, where quadratization, embedding overhead, or multiple qbsolv subcalls become dominant and time-consuming. 

\begin{table}[H]
\centering
\renewcommand{\arraystretch}{1.2}

\begin{tabularx}{\textwidth}{|Y|L|M|N|W|}
\hline
{Molecule} & {Method} & {Hardware} & {Reported / Estimated TTS} & {Note / Source}\\
\hline
H$_2$ &
VQE \cite{Bentellis2023} &
IBMQ Manila &
$334 \pm 40$ s &
End-to-end TTS includes queue, compilation, execution, and readout.\\
\hline
H$_2$ &
VQE \cite{Bentellis2023} &
AQT Marmot &
$13{,}297 \pm 625$ s &
End-to-end TTS includes queue, compilation, execution, and readout.\\
\hline
H$_2$O &
VQE \cite{Jones2024} &
IBMQ Kolkata &
$5,400$ s &
Quantum Computation Time.\\
\hline
H$_2$O &
XBK \cite{Copenhaver2021} &
D-Wave &
more than 8 s (est.) &
Estimated from Fig.~3 in the paper (time vs. pre-quadratization qubits).\\
\hline
H$_2$O &
QCC \cite{Copenhaver2021} &
D-Wave &
more than 6 s (est.) &
Estimated from Fig.~3.\\
\hline
{H$_2$} &
{Quantum-Inspired Ising Machine} &
{GPU} &
{1.2 s} &
{Proposed efficient simulation using the Ising Machine approach.}\\
\hline
{H$_2$O} &
{Quantum-Inspired Ising Machine} &
{GPU} &
{2.4 s} &
{Proposed efficient simulation using the Ising Machine approach.}\\
\hline
\end{tabularx}

\caption{Comparison of TTS between quantum hardware and quantum-inspired simulations for solving the electronic structure in small molecular systems. Notably, the quantum-inspired algorithms compute the entire energy profile for 
$H_2$ and $H_2O$
 in 1.2 s and 2.4 s, respectively, whereas quantum hardware typically requires over 6 s to compute just a single energy point with comparable error.}
\label{tab:tts_comparison}
\end{table}

In summary, the numerical results presented here demonstrate the effectiveness of quantum-inspired CIM and SB algorithms in computing molecular ground-state energies. The algorithms achieve promising accuracy for H$_{2}$ across a range of conditions, indicating their potential scalability to more complex systems such as H$_{2}$O. These findings highlight the practicality of CIM- and SB-based approaches as alternative strategies for electronic structure calculations in the NISQ era and beyond.

\section{Conclusion}\label{sec:conclusion}

In this work, we studied the potential of quantum-inspired algorithms, specifically variants of CIMs and SB, for computing molecular ground-state energies. Using hydrogen ($H_2$) and water ($H_2O$) as representative test cases, we demonstrated that these approaches not only reproduce key features of molecular dissociation curves but also produce energy profiles consistent with the expected physical behavior. For $H_2$, the CFC algorithm, as the most powerful variant of CIMs, achieved high accuracy in a wide range of parameters, while the extension to $H_2O$ illustrated their scalability in more complex molecular systems.

Our numerical analysis, consistent with recent studies that compare the performance of QAIA \cite{Zeng2024}, indicates that CFC variant exhibits superior efficiency compared to other CIM-inspired algorithms \cite{Zeng2024,Leleu2019}.

Compared with conventional optimization strategies, the CIM and SB algorithms offer several distinct advantages. Their foundations in nonlinear dynamics and bifurcation theory, combined with a mean-field treatment of the quantum behavior of each pulse, enable efficient exploration of complex energy landscapes. The collective and superposed evolution of spin states allows the system to simultaneously probe multiple solution pathways, increasing the likelihood of identifying low-energy configurations with relatively modest computational resources. Moreover, combining these methods with classical post-processing techniques, such as steepest descent, yields a hybrid framework that enhances accuracy while reducing the sampling overhead often required in stochastic approaches. Together, these features establish CIM and SB as compelling alternatives to classical electronic-structure methods and quantum annealing-based platforms, particularly in the NISQ era.

It is noteworthy that our proposed GPU-based algorithm for quantum chemistry calculations can efficiently generate large numbers of samples, enabling the treatment of large-scale scientific problems with minimal time and resource overhead due to its inherent parallelism \cite{Reifenstein2021,Lvovsky2019,Goto2019,Goto2021}. This high-throughput sampling not only facilitates exploration of Ising-form problems but also provides a powerful framework for investigating electron configurations in molecular systems to identify their stable states. With this physics-inspired, GPU-accelerated approach, the exploration of electronic structures in complex molecules can be performed more efficiently, offering promising applications in material design—such as optimizing catalytic surfaces, studying superconducting materials, and engineering novel semiconductors—and in drug discovery, for instance, in predicting binding affinities, screening molecular conformations, and modeling reaction pathways in biomolecular systems.

Most importantly, in contrast to purely digital optimization approaches, the physical realization of CIMs offers an analog computational paradigm with distinct advantages for quantum chemistry applications. CIMs evolve continuous dynamical variables, such as optical field amplitudes, thereby avoiding the numerical precision limitations and significant memory overhead associated with high-accuracy digital computation. Furthermore, intrinsic quantum noise in optical parametric oscillator networks provides a physically meaningful source of stochasticity, enabling efficient exploration of complex energy landscapes and reducing the likelihood of stagnation in local minima \cite{Kumagai2025}. Although the present study employed SimCIM, it is noteworthy that hardware-realized CIMs are expected to offer additional advantages, including faster physical convergence, enhanced scalability, and improved sampling fidelity due to genuine quantum fluctuations. When combined with quantum-inspired algorithms such as CFC,  SFC, CAC, and dSB variants as demonstrated in this work, CIM-based hybrid solvers present a scalable and physics-informed alternative to both classical heuristics and hardware-restricted quantum annealers. Their demonstrated ability to approximate molecular ground-state energies using modest computational resources highlights their potential as a key component of future analog–digital hybrid computing frameworks in quantum chemistry.

During the preparation of this work, Takesue et al. \cite{Takesue2025} demonstrated that a CIM based on degenerate optical parametric oscillators can efficiently solve classically hard NP problems, such as finding maximum independent sets in large graphs, by mapping them onto an Ising Hamiltonian and exploiting the machine’s intrinsic parallelism and energy minimization dynamics. This success story provides a testimony that photonic-based, quantum-inspired systems can efficiently explore complex energy landscapes, a capability directly relevant to quantum chemistry, where molecular Hamiltonians can be encoded into Ising-type models to determine ground-state energies. Leveraging CIMs thus enables systematic exploration of electronic configurations, outperforming traditional digital algorithms in variational molecular simulations.

Building on this foundation, future research can enhance convergence and accuracy through optimized parameter tuning, noise mitigation, and adaptive sampling, while integration with machine learning may enable automated optimization. Embedding these methods in hybrid quantum–classical workflows could further improve efficiency. Extending the approach to larger, chemically relevant systems, including complex biomolecules, remains an important opportunity, and the hybrid algorithm could also be adapted to compute excited-state energy profiles \cite{González2012}, broadening its applicability in quantum chemistry, materials science, and drug discovery.

\section*{Author Contributions}
M. H. and H. S. contributed equally to this work, and N. A. A. supervised the project.
\section*{Code and Data Availability}
The codes and data for supporting this paper is available at 
\\ https://github.com/MahmoodSpewAfsh/Quantum-Inspired-Algorithms-for-Quantum-Chemistry-Calculations


\end{document}